\documentclass[10pt,conference,letterpaper]{IEEEtran}
\usepackage{times,amsmath,epsfig}
\usepackage{amssymb}
\usepackage{graphicx}
  \usepackage{color}
   \usepackage{url} 
    \usepackage{float}
    \usepackage{wrapfig}

\usepackage{subfigure}
\title{Semantic-driven Configuration of Internet of Things Middleware}

\author{
{Charith Perera{\small $~^{*\#1}$}, Arkady Zaslavsky{\small $~^{*3}$}, Michael Compton{\small $~^{*3}$}, Peter Christen{\small $~^{\#4}$}, Dimitrios Georgakopoulos{\small $~^{*5}$} }
\vspace{1.6mm}\\
\fontsize{10}{10}\selectfont\itshape
$^{\#}$Research School of Computer Science, The Australian National University, Canberra, ACT 0200, Australia\\
\fontsize{9}{9}\selectfont\ttfamily\upshape
$^{4}$peter.christen@anu.edu.au
\fontsize{9}{9}\selectfont\ttfamily\upshape
\vspace{1.2mm}\\
\fontsize{10}{10}\selectfont\rmfamily\itshape
$^{*}$CSIRO ICT Center, Canberra, ACT 2601, Australia\\
\fontsize{9}{9}\selectfont\ttfamily\upshape
\{$^{1}$charith.perera,
$^{2}$michael.compton,
$^{3}$arkady.zaslavsk,
$^{5}$dimitrios.georgakopoulos\}@csiro.au
}

\begin{document}
\maketitle
\begin{abstract} 
We are currently observing emerging solutions to enable the Internet of Things (IoT). Efficient and feature rich IoT middeware platforms are key enablers for IoT. However, due to complexity, most of these middleware platforms are designed to be used by IT experts. In this paper, we propose a semantics-driven model that allows non-IT experts (e.g. plant scientist, city planner) to configure IoT middleware components easier and faster. Such tools allow them to retrieve the data they want without knowing the underlying technical details of the sensors and the data processing components. We propose a \textit{Context Aware Sensor Configuration Model} (CASCoM) to address the challenge of automated context-aware configuration of filtering, fusion, and reasoning mechanisms in IoT middleware according to the problems at hand. We incorporate semantic technologies in solving the above challenges. We demonstrate the feasibility and the scalability of our approach through a prototype implementation based on an IoT middleware called Global Sensor Networks (GSN), though our model can be generalized into any other middleware platform. We evaluate CASCoM in agriculture domain and measure both performance in terms of usability and computational complexity.
\end{abstract}

\section{Introduction}
\label{sec:Introduction}

The Internet of Things (IoT), an emerging paradigm, provides a networked infrastructure that enables \textit{things} to be connected anytime, anyplace, with anything and anyone, ideally using any path, any network and any service \cite{P029}. The \textit{things} in IoT are accompanied with sensors and actuators. It is estimated that there are about 1.5 billion Internet-enabled PCs and over 1 billion Internet-enabled mobile phones today. By 2020, there will be 50 to 100 billion devices\footnote{We use terms \textit{objects}, \textit{things}, \textit{smart objects}, \textit{devices}, \textit{nodes} to give the same meaning as they are frequently used in IoT related documentation interchangeably.} connected to the Internet \cite{P029}. Since these smart devices comprise sensors, it is evident that there would be many sensors deployed around us in the future. Even today, sensors are used in many domains such as agriculture, environmental monitoring \cite{ZMP007}. In order to analyse and understand a given phenomenon extensively, data generated from appropriate sensors need to be fed into more sophisticated applications. These applications are designed to produce certain results once they are given required sensor data as inputs. IoT middleware solutions help to retrieve data from sensors and feed them into applications easily by acting as a mediator between the hardware layer and the application layer. In order to perform such a binging act,  middleware solutions need to be configured themselves depending on the context information and user requirements. Our objective is to automate and simplify the configuration of an IoT middleware and improve its usability so non-IT experts can use it efficiently and effectively. Our contribution can be listed as follows:
\vspace{-5pt}
\begin{itemize}
\item We develop a configuration model called CASCoM to enrich an existing IoT middleware. This model helps non-IT experts to configure sensors and data processing components using a \textit{single-click} quickly and easily. As the final outcome, CASCoM produces data streams that can be fed into applications/services easily where further processing may occur.

\item Our model automates the configuration process which is a significant improvement over the  current Global Sensor Network (GSN) \cite{P022} approach where all the configurations need to be done manually by IT experts.

\item CASCoM is completely driven by semantic annotated data at the back end. Therefore, new sensors and data processing components can be added at any time. No changes are required in the application.


\item CASCoM allows the users to discover additional context information.

\item We provide a cost calculation model that considers and combines software and hardware costs when configuring sensors and data processing components. It also allows  users to define their own priorities.

\item CASCoM is capable of suggesting and advising future sensor deployments, if the existing sensors are incapable of fulfilling user requirement.

\end{itemize} 

The remainder of this paper is organized as follows. In Section \ref{sec:Background}, we describe the background and motivation behind our work. The problem we addressed in this paper is comprehensively analysed and presented with use-case scenarios in Section \ref{sec:Problem_Definition}. In Section \ref{sec:Proposed_Solution}, we propose our solution, CASCoM, in detail. Implementation details are explained in Section \ref{sec:Implementation}. In Section \ref{sec:Discussion}, we evaluate the CASCoM in both qualitative and quantitative methods. Related work are reviewed in Section \ref{sec:Related_Work}. Finally, we present the conclusions and future directions.

\section{Background}
\label{sec:Background}

 Broadly, configuration in IoT paradigm can be categorized into two: \textit{sensor-level} configuration and \textit{system-level} configuration. Sensor-level configuration focuses on changing a sensor's behaviour by configuring  its embedded software parameters such as sensing schedule, sampling rate, data communication frequency, communication patterns and protocols. In this paper, we are focused on developing a system-level configuration model for IoT midddleware platforms. Specifically, our proposed model identifies, composes, and configures both sensors and data processing components according to the user requirements.

 The challenge of configuring an IoT middleware solution can be understood by analysing an existing middleware such as Global Sensor Networks (GSN) \cite{P022}. The high-level architecture and the data flow diagram of the GSN middleware is presented in \cite{P022}. \textit{Wrappers} perform the hardware-level communication with  sensors. Each and every sensor that needs to be connected to the GSN middleware should have a corresponding \textit{wrapper}. In order to retrieve sensor data, users are required to define their requirements using a XML file called \textit{Virtual Sensor Definition (VSD)} \cite{P022}. Once the \textit{wrapper} receives data, it forwards them to the \textit{Virtual Sensor (VS)} as specified in the VSD. Similarly, multiple wrapper may send data to a single VS. A \textit{VS} may have any number of input data streams and produces exactly one output data stream. In GSN, sensor data can be processed in three layers: (1)\textit{virtual sensors layer}, (2) \textit{query processing layer}, and (3) \textit{applications and services layer} (outside GSN middelware). 
 
 The query processing layer can perform filtering and integration tasks based on SQL-like specifications. However, data processing tasks that cannot be accomplished using SQL need to be performed either in layer 1 or 3. Layer 3 consists of sophisticated applications (and services) that take  specific data streams and perform complex data processing operations. For example, an application may take air temperature, air humidity, and leaf wetness as the input data stream. Then, it generates a map by visualizing how a certain type of disease may spread across an agricultural field. Such complex data processing and modelling tasks are out of the scope of GSN's processing capabilities. The responsibility of an IoT middleware (such as GSN) is to generate the appropriate data streams (that the applications require as inputs) without (or with minimum) user intervention. In order to accomplish this, layer 1, which we focus in this paper, needs to play a critical role. The virtual sensors layer allows to apply data processing operations (less complex operations compared to layer 3) over the sensor data. In the existing  GSN, all the data processing components  in layer 1 need to be developed by the user and need to be manually selected based on the user requirements. Performing such task manually is tedious and cumbersome for non-IT experts (such as plant scientists and environmentalists). 
 
 Let us discuss the term \textit{data processing components} in relation to layer 1. Data processing can be defined as manipulation of input data with an application program to obtain desired output. In layer 1, data processing components perform operations such as filtering, fusing, reasoning, anomaly detection, unit conversion, missing value estimation, noise reduction, feature extraction and so on. Operations should be able to be performed within acceptable time frame (i.e. in real-time, ideally before the next data packets arrive) as we are dealing with data streams. In order to fulfil the user requirements, several data processing components\footnote{We use the term \textit{components}, but can be called \textit{functions}, \textit{methods}, and \textit{modules}.}  may be required to compose together. 

 In addition to the manual configuration activities that need to be performed, there are several weaknesses in the current approach. Figure \ref{Figure:configuration_Workflow_Comparison}(a) illustrates the activity diagram of the existing configuration work-flow of the GSN middleware. 
\begin{itemize}
\item Users need to know the low-level details such as data types and measurement units of the sensors in order to define the VSD manually. 

\item It is extremely difficult to memorise different combinations of sensor types that can be used to fulfil user requirements (which sensors need to be composed together to detect an event?). 


\item  Users need to know the availability of data processing components, their input/output data types and their capabilities  to develop a strategy. Data processing operations need to be applied on data in the correct sequence. 

\item  There is no way to find out the strategies to overcome the issues when existing hardware resources (i.e. existing sensors) and software resources (i.e. data processing components) are incapable of producing the results that users require. 

\item  Further, the solutions designed by users may not be the optimum solution (e.g. due to the variability of hardware and software costs).

\end{itemize}

In existing GSN middleware, many configuration files and programming codes need to be manually defined by the users (without any help from GSN). An ideal IoT middleware configuration model should address all the above mentioned challenges. The configuration model we propose in this paper is applicable towards several other emerging paradigms, such as sensing as a service \cite{ZMP003}.

\section{Problem Analysis}
\label{sec:Problem_Definition}

\begin{figure}[b!]
 \centering
\vspace{-0.63cm}
 \includegraphics[scale=.75]{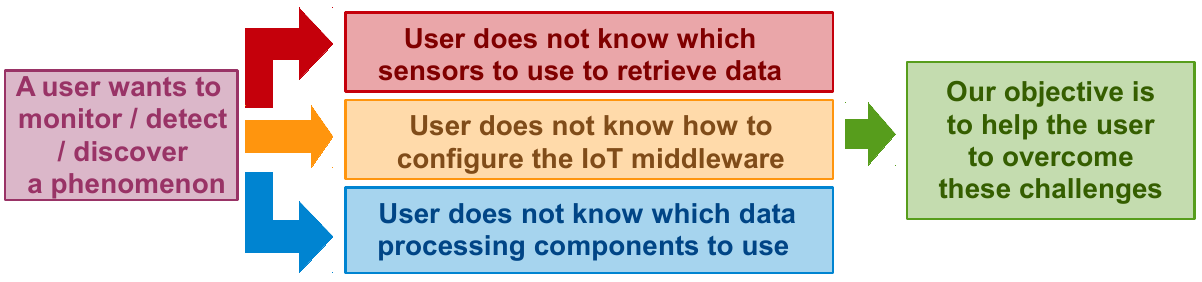}
\vspace{-0.35cm}	
 \caption{The problem definition in general}
 \label{Figure:Problem_Analysis_General}	
\end{figure}

This section describes and analyses the problem we address in this paper with concrete examples and scenarios. Figure \ref{Figure:Problem_Analysis_General} illustrates the problem in general. The explanations are based on agriculture and environmental monitoring domains. The proposed solution helps users to overcome difficulties listed above. Our research question is \textit{`How to develop a model that allows non-IT experts to configure sensors and data processing mechanisms in an IoT middleware according to the user requirements?'}. Extended explanations are provided in \cite{ZMP004}.

 Let us introduce the notations that we are going to use in this paper: Sensor ($S^{t}$) where $t$=type (e.g. each sensor type is represented by a different number: 1=air temperature, 2=air humidity, 3=leaf wetness, 4=Carbon Monoxide, 5=Carbon Dioxide, 6=Molecular Oxygen, 7=Methane, 8=Nitrogen Dioxide); Wrapper ($W$); Virtual Sensor ($VS$); data processing component ($C_{id}^{f}$) where $f$=function (e.g. each function is represented using a unique number. 1=\textit{airStressDetector}, 2=\textit{phytophtoraMonitor}, 3=\textit{pollutionDetector}) and $id$ = developer unique identifier.

\begin{figure}[b!]
 \centering
\vspace{-0.43cm}
 \includegraphics[scale=.72]{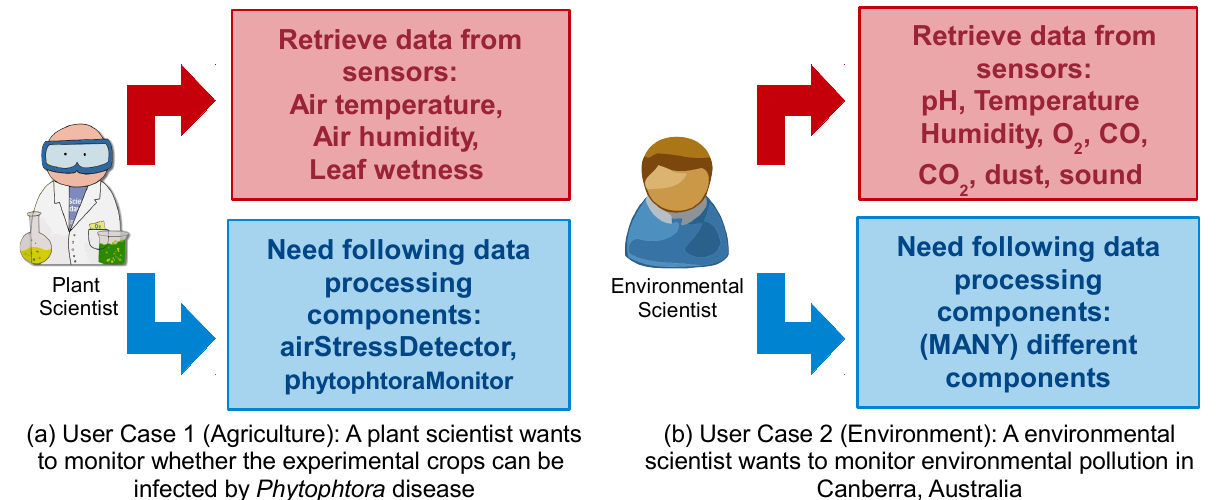}
\vspace{-0.43cm}	
 \caption{Use cases that illustrates the need of CASCoM}
 \label{Figure:Problem_Analysis}	
\end{figure}

Figure \ref{Figure:Problem_Analysis} illustrates two scenarios from two different domains. Each of them has different user requirements that lead to two different execution flows. We selected these two scenario due to the fact that, together, they allow us to showcase the full capabilities of CASCoM. In use case 1, a plant scientist wants to monitor whether the experimental crops can be infected by \textit{Phytophtora} \cite{P452} disease or not. \textit{Phytophtora} is a fungal disease which can enter a field through a variety of sources. The development and associated attack of the crop depends strongly on the climatological conditions within the field. Humidity plays a major role in the development of \textit{Phytophtora}. Both temperature and whether or not the leaves are wet are also important indicators to monitor \textit{Phytophtora}. The following facts explain \textit{Phytophtora} monitoring (simplified for demonstration purposes). It is important to highlight that rule-based reasoning does not intended to replace rule engines \cite{P598}. The objective here is to create the data items that are required by the application.

\begin{itemize}
\footnotesize
 \item IF \textbf{airTemperature}  $\textless$ $\alpha$ AND \textbf{airHumidity} $\textless$ $\beta$ THEN \textbf{airStress} level = low ELSE \textbf{airStress} level = high

 \item IF \textbf{airStress} = high  AND \textbf{leafWetness} $\textgreater$ $\delta$ THEN \textit{PhytophtoraDisease} = Can-be-infected ELSE = Cannot-be-infected

\end{itemize}

One of the responsibility of an IoT middleware is to combine different sensors and data processing components autonomously and produce a data stream. A user can feed the data stream into an application for further complex processing such as visualization and modelling that allows the user to achieve their objectives. The main challenge is that the plant scientist may not know (or remember) the above facts (rules). Further, we should not expect a plant scientist to write XML or Java code as part of the configuration. An ideal IoT middleware should help the scientist (non-IT expert) to overcome these challenges by providing tools that are easy to use. The scientist should be able to configure the middleware according to the problems/tasks at hand with minimum effort. Additionally, advanced customization will be useful to optimize the configuration process. Comparatively, use case 1 is less complex as there is only one way to monitor the disease (above rules). For example, the sensor types and data processing components need to be used are straight forward. 

\begin{itemize}
\footnotesize

 \item \textbf{Use case (1) Solution:} $   \left (( S^{1}, S^{2} )  \Rightarrow  C_{1}^{1}, S^{3} \right )\Rightarrow C_{1}^{2}  $
 
 

\end{itemize}

Configuration becomes a complex task in the use case 2. In this scenario, an environmental scientist wants to measure the environmental pollution in Canberra, Australia. In comparison to the use case 1, there are many different ways to measure and visualize pollution. Different sensors and data processing components can be combined together to fulfil the user requirements as listed below.


\begin{itemize}
\footnotesize

  \item \textbf{Use case (2) Solution 1:} $   \left ( S^{4}, S^{5} , S^{6} , S^{7} , S^{8}\right ) \Rightarrow  C_{38}^{3}  $

  \item \textbf{Use case (2) Solution 2:} $   \left ( S^{5}, S^{8}  \right ) \Rightarrow  C_{77}^{3}  $

  \item \textbf{Use case (2) Solution 3:} $  \left ( S^{1}, S^{5}, S^{7} \right ) \Rightarrow  C_{32}^{3}  $

\end{itemize}

In such circumstances, it is important to consider context information (e.g. accuracy, reliability) and cost of data acquisition (e.g. data communication time and computation time). This allows a user to make the final decision on which solution to be used depending on the cost and context factors. Both hardware and software costs need to be considered. Additionally, users may need to discover additional context information \cite{ZMP007}. Depending on the user requirements and layer 3 application requirements, the required output data stream may vary. Sample outputs, in relation to use case 1, are listed below.

\begin{itemize}
\footnotesize

  \item \textbf{Output 1:} airTemperature [double], airHumidity [double], airStress [string], \\leafWetness [double], PhytophtoraDisease [boolean]
  
  \item \textbf{Output 2:} PhytophtoraDisease [boolean], location [string], batteryLevel[double]

\end{itemize}

Finally, when existing resources are insufficient to satisfy the user requirements, it is very important to advice the users regarding possible improvements.

\section{The CASCoM Architecture}
\label{sec:Proposed_Solution}

\begin{figure}[h]
 \centering
 \includegraphics[scale=.70]{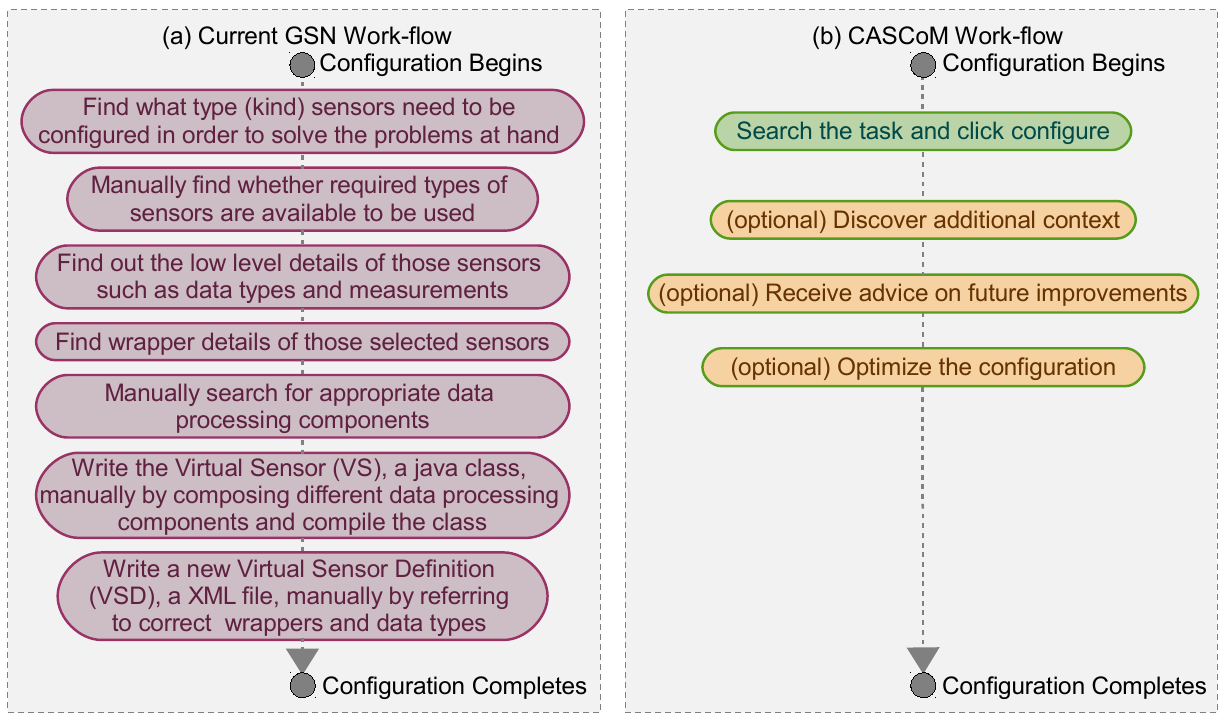}
\vspace{-0.43cm}	
 \caption{Configuration Execution-flow Comparison: (a) Current GSN  (b) CASCoM}
 \label{Figure:configuration_Workflow_Comparison}	
\vspace{-0.63cm}	
\end{figure}

\begin{figure}[b]
 \centering
 \vspace{-0.63cm}
 \includegraphics[scale=.68]{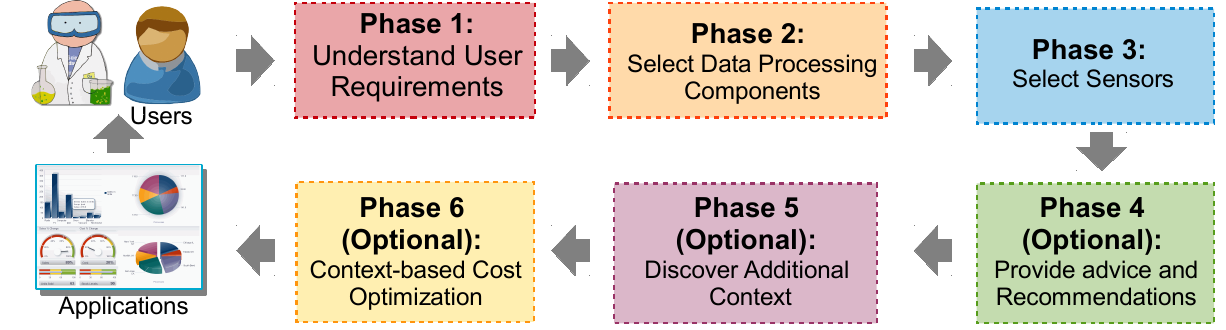}
 \caption{The Context-Aware Sensor Configuration Model (CASCoM)}
 \label{Figure:The_Model}	
\end{figure}

Based on the challenges we identified in Section \ref{sec:Problem_Definition}, we designed a model, which is supported by a tool, to overcome the difficulties. Context-Aware Sensor Configuration Model (CASCoM) simplifies the IoT middleware configuration process significantly. Figure \ref{Figure:configuration_Workflow_Comparison} compares the execution-flow of sensor configuration in the current GSN approach and the CASCoM approach. As it is clearly visible, the current GSN model requires a number of steps to be executed by IT experts. In contrast, our proposed model allows non-IT experts to configure IoT middleware using a \textit{single-click}. All the difficulties are handled internally behind the scene without the user involvement. Additionally, we offer several advance features that allow optimization and customization. As depicted in Figure \ref{Figure:The_Model}, CASCoM consists of six phases. Some phases may or may not be visible to the users. Phases are different from the steps needed to be followed in the CASCoM Tool.

\textbf{CASCoM Execution Flow:} In phase 1, users are facilitated with a graphical user interface, which is based on a \textit{question-answer (QA)} approach, that allows to express the user requirements. Users can answer as many question as possible. CASCoM searches and filters the tasks that the user may want to perform. From the filtered list, users can select the desired task. The details of the QA approach are presented later in this section. In phase 2, CASCoM searches for different programming components that allow to generate the data stream required. In phase 3, CASCoM tries to find the sensors that can be used to produce the inputs required by the selected data processing components. If CASCoM fails to produce the data streams required by the users due to insufficient resources (i.e. unavailability of the sensors), it will provide advice and recommendations on future sensor deployments in phase 4. Phase 5 allows the users to capture additional context information. The additional context information that can be derived using available resources and knowledge are listed to be selected.  In phase 6, users are provided with one or more solutions\footnote{Solution is a combination of sensors and data processing components that can be composed together in order to satisfy the user requirements.}. CASCoM calculates the costs for each solution. By default, CASCoM will select the solution with lowest cost. However, users can select the cost models (discussed later in this section) as they required. Finally, CASCoM generates all the configuration files and program codes which we listed in Figure \ref{Figure:configuration_Workflow_Comparison}(a). Data starts streaming soon after.

\begin{figure}[b!]
 \centering
 \vspace{-0.93cm}
 \includegraphics[scale=.70]{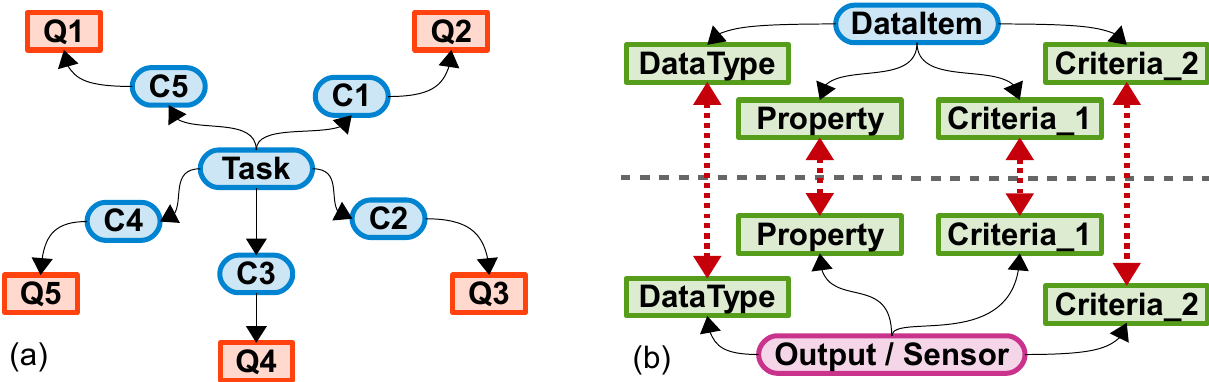}
\vspace{-0.43cm}	
 \caption{(a) A part of QA-TDO shows how we developed the QA model. It is important to note the pattern (i.e. \textit{Task} $\rightarrow$ \textit{Concept} $\rightarrow$ \textit{Question}).  (b) shows how validation can be performed using semantic data.}
 \label{Figure:Valiication}	
\end{figure}

\begin{figure}[t]
 \centering
 \includegraphics[scale=.56]{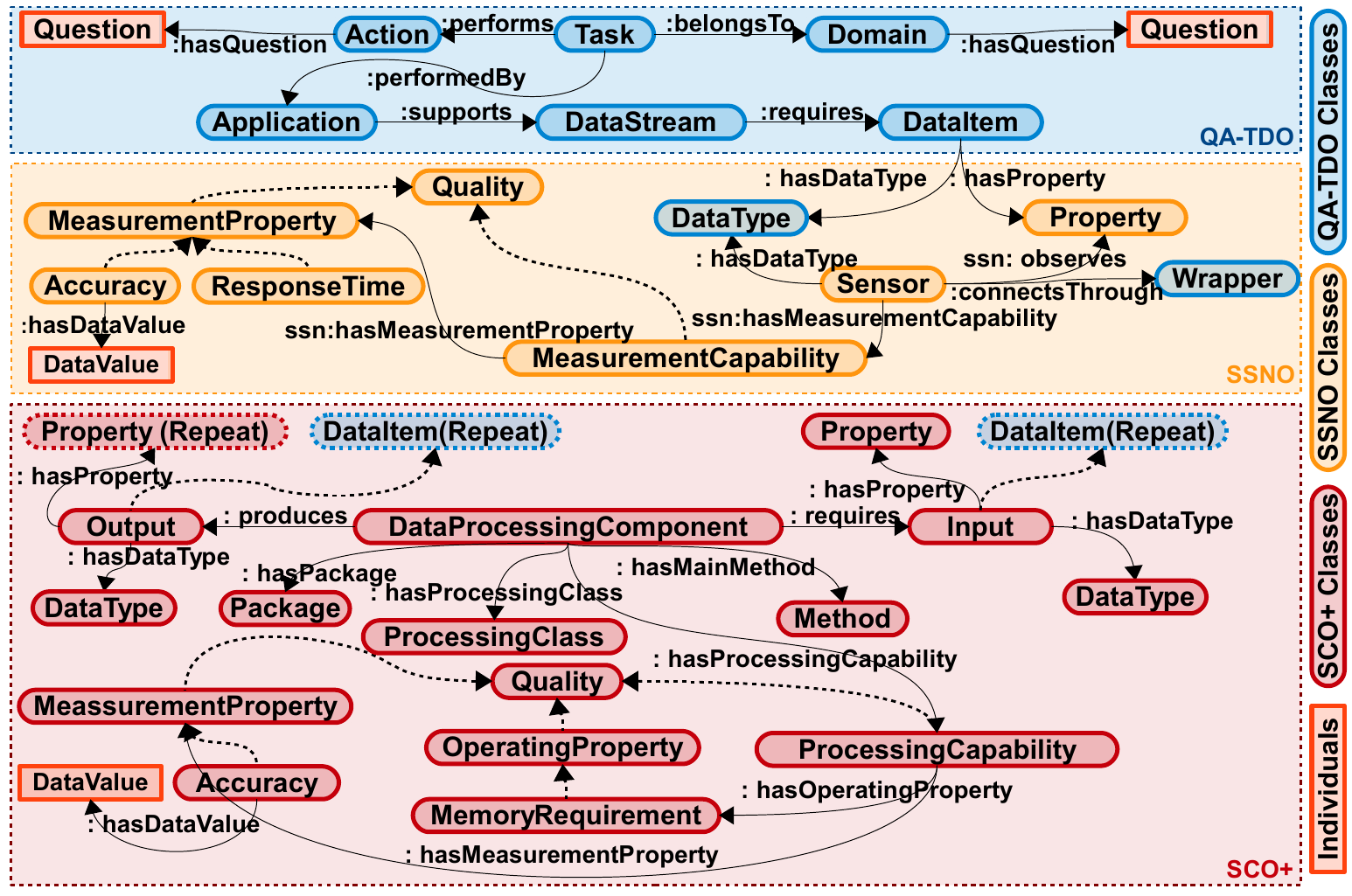}
\vspace{-0.43cm}	
 \caption{Extracts of different ontological data models used in CASCoM: QA-TDO, SCO \cite{P612}, and SSNO \cite{P626}. The colour coding refers to different prefixes.}
 \label{Figure:Ontology_Model}	
\vspace{-0.73cm}	
\end{figure}

\textbf{Phase 1: Understand User Requirements:} The objective of this phase is to help  users to search for a task that they need to perform easily from a large number of possibilities. For example, users are allowed to narrow down the possibilities by mentioning facts such as domain (e.g. agriculture), and type of the task (e.g. event, visualization). In order to increase the usability, CASCoM retrieves the facts from the users through a QA model (Sample questions: Do you want to visualize data?, Do you want to detect an event?, Do you want to monitor a disease infection? What is the domain your task is related to?). When a user answer a question, the remaining questions will be dynamically selected based on the previous answer. An extract of the proposed \textit{Question and Answer oriented Task Description Ontology} (QA+TDO) is presented in Figure \ref{Figure:Ontology_Model}. In QA+TDO, tasks can be explained by any concept as depicted in $C1$, $C2$, etc. in Figure \ref{Figure:Valiication}(a). Each concept should have a \textit{`hasQuestion'} property which links to a question (i.e. $Q1$, $Q2$ and so on). In QA+TDO, $C$  are answers to the questions. (e.g. If $Q1$= What is the domain your task is related to?, then $C5$ is \textit{`domain'} and an individual of $C5$ can be \textit{`agriculture'}.). The extensibility and scalability of this approach is discussed in Section \ref{sec:Discussion}.

\textbf{Phase 2 and 3: Select Sensors and Data Processing Components:} CASCoM requires all the information related to sensors and data processing components to be stored in a repository. We extended the Software Component Ontology \cite{P612} (SCO) as presented in Figure \ref{Figure:Ontology_Model} in order to model information about data processing components. Further, we modelled sensor descriptions using semantic Sensor Ontology (SSNO) \cite{P626}. In this phase, the software components are selected in such as a way that they can together produce the data stream required to perform the task selected in phase 1. For example, in order to monitor \textit{PhytophtoraDisease}, first CASCoM searches for a software components that can be used to produce the required data. It first finds  \textit{PhytophtoraDisease Detector}.  The inputs it requires are \textit{air stress} and \textit{leaf wetness}. The phase 3 selects the sensors that produce the output that matches the inputs of the selected component. \textit{Leaf wetness} can be measured directly using hardware sensors. However, \textit{air stress} cannot be detected using any physical sensor. This requires CASCoM to execute phase 2 again in order to find a software component that produces \textit{air stress}. Then CASCoM finds  \textit{Air Stress Detector} which takes \textit{air temperature} and \textit{air humidity} as inputs and produces \textit{air stress} as the output. Further, \textit{air temperature} and \textit{air humidity} can be sensed directly through hardware sensors. The IoT middleware configuration process will be completed once the required sensors and data processing components are identified. The remaining phases are optional. CASCoM performs validation as illustrated in Figure \ref{Figure:Valiication}(b). During the sensors and data processing components composition process, different criteria are evaluated  (e.g. data types: int, boolean / measurement units: Celsius, Fahrenheit) in order to verify whether the inputs and outputs are compatible.

\textbf{Phase 4 (Optional): Provide Advice and Recommendations:} Through comparing SSNO and SCO, this phase identities the resource insufficiencies and provide advice to the users regrading future sensor deployments and software component acquisition. This phase provide alternative advices if there are multiple ways to address the insufficiencies based on the solution (e.g. use case 2).

\textbf{Phase 5 (Optional): Additional Context Discovery:} With the help of knowledge modelled in ontologies, this phase discovers context information that can be derived by using sensor data. \textit{`Context is any information that can be used to characterise the situation of an entity. An entity is a person, place, or object that is considered relevant to the interaction between a user and an application, including the user and applications themselves'} \cite{ZMP007}. Additional context information such as sensor location and sensor battery life may be required by applications in order to perform complex tasks such as geographical based visualization and developing energy-aware sensing schedules. Therefore, discovering additional context as such is important. Each application may have a compulsory set of inputs that it needs to perform the primary task, though they may accept additional context information in order to provide enhanced results.

\textbf{Phase 6 (Optional): Context-based Cost Calculation:} CASCoM performs ontological reasoning to find out all possible solutions. Each solution may combine different sensors and data processing components where their costs may different. For example, different types of sensors can be used to monitor environmental pollution (refer Section \ref{sec:Problem_Definition}). In CASCoM, cost does not always refer to financial terms (e.g. sensors: energy, bandwidth, latency; data processing: memory requirement, processing time). By default, all the context parameters are treated equally. However, users can define their priorities for each context property in comparative fashion \cite{ZMP006}. If the  users want more \textit{reliable} sensors, the \textit{reliability} can be defined with more priority, but it may increase the cost.

\section{Implementation}
\label{sec:Implementation}
\vspace{-5pt}

\begin{figure*}
 \centering
\vspace{-0.53cm}
 \includegraphics[scale=0.99]{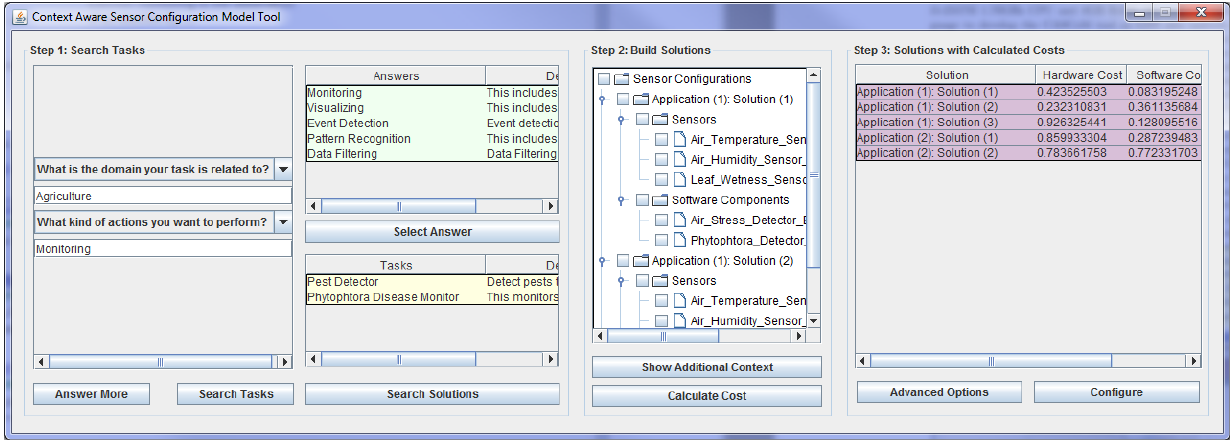}
 \vspace{-0.23cm}	
 \caption{User interface that supports CASCoM}
 \label{Figure:User Interface}	
\vspace{-0.55cm}	
\end{figure*}

This section presents programming level details of our proof of concept development and evaluation. Hardware and software platforms, APIs and frameworks, semantic data models, and sample data sets we used to evaluate the performance of the prototype implementation are explained in this section. For proof of concept deployment and evaluation, we used a computer with Intel(R) Core i5-2557M 1.70GHz CPU and 4GB RAM. We used the Java programming language to develop the CASCoM tool as GSN also natively supports Java. We employed the open source Apache Jena API (jena.apache.org) to manipulate semantic data. In addition, we used the Apache Commons mathematics (commons.apache.org/math) library for advanced cost calculations based on user priorities. The costs are calculated using a weighted Euclidean distance-based indexing technique called \textit{Comparative Priority-based Weighted Index} (CPWI)\cite{ZMP006}.

We modelled 40 sensor descriptions according to the \textit{Semantic Sensor Network Ontology} (SSNO) \cite{P626}. Additional extensions are added to SSNO in order to model context information related to sensors as explained in \cite{ZMP006}. Further, we modelled 40 data processing component descriptions according to the Software Component Ontology Plus  (SCO+). SCO+ is based on SCO \cite{P612}, but additionally  supports modelling context information such as execution time and reliability as presented in Figure \ref{Figure:Ontology_Model}. We modelled context information in SCO+ using an approach similar to SSNO. The data processing components may take any number of inputs and produce one outputs. We employed our previous work, CASSARAM \cite{ZMP006} to search sensors based on context properties and to calculate costs. As a result of the integration of CASCoM into the GSN middleware, virtual sensors and  virtual sensor definitions were generated  autonomously. All the other GSN components remained same. We introduced several new components under 4 different managers: [QA Manager] \textit{QA Filter}, \textit{Dynamic SPARQL Generator}; [Task Manager] \textit{Solutions Finder}, \textit{Solution Composer}, \textit{Solution Validator}; [Services Manager] \textit{Cost Calculator}, \textit{Context Discovery Manager}, \textit{Solution Adviser}; [Configuration Manager] \textit{VS Generator}, \textit{VSD Generator}, \textit{Wrapper Handler}. The user interface of the CASCoM tool is presented in Figure \ref{Figure:User Interface}. It is a critical component of the proposed model as it significantly help the users to configure the IoT  middleware easier and faster.

\section{Evaluation, Discussion and Lessons Learned}
\label{sec:Discussion}

\begin{figure}[t]
\centering
\mbox{\subfigure{\includegraphics[scale=0.27]{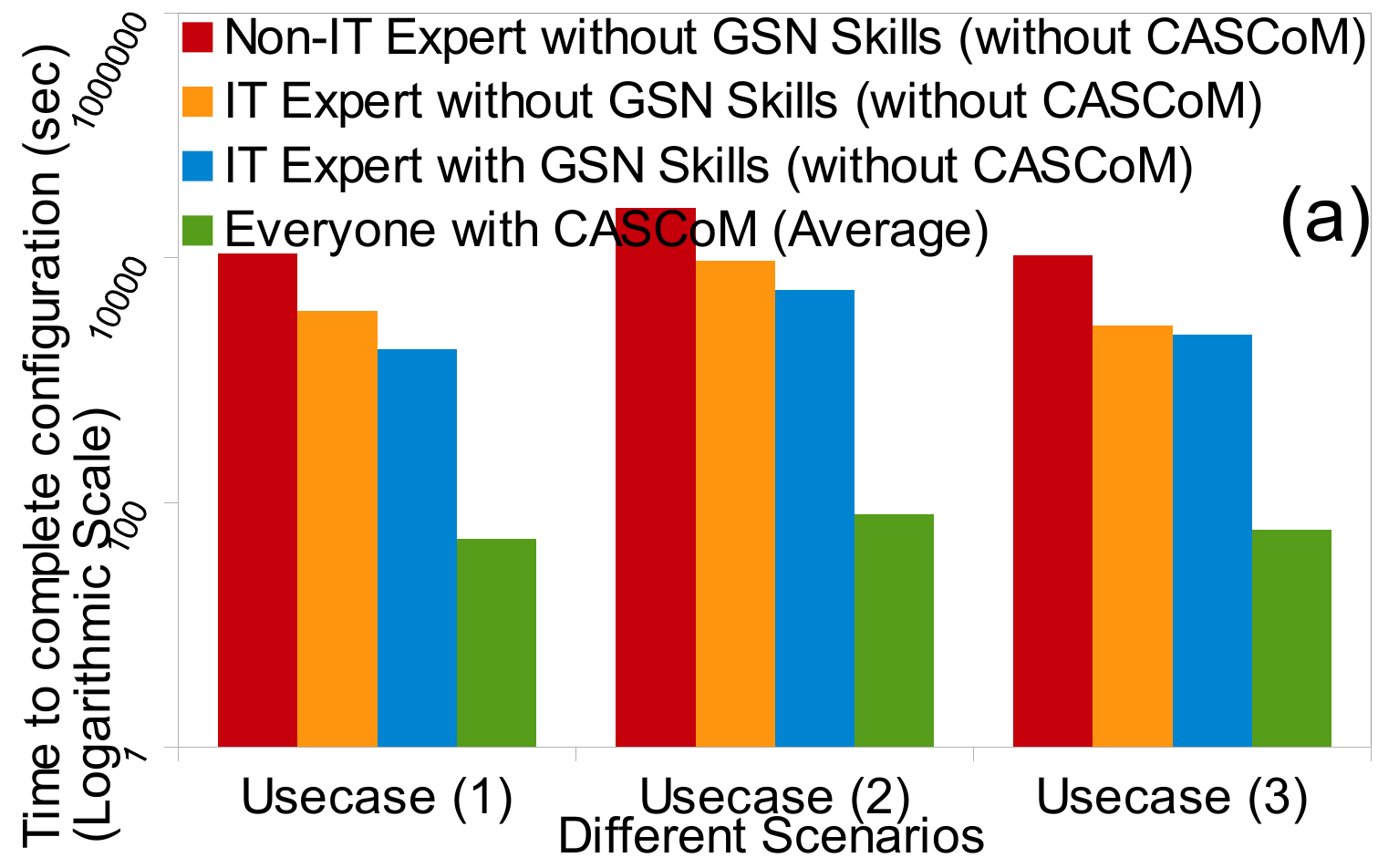}
\quad
\subfigure{ \includegraphics[scale=0.27]{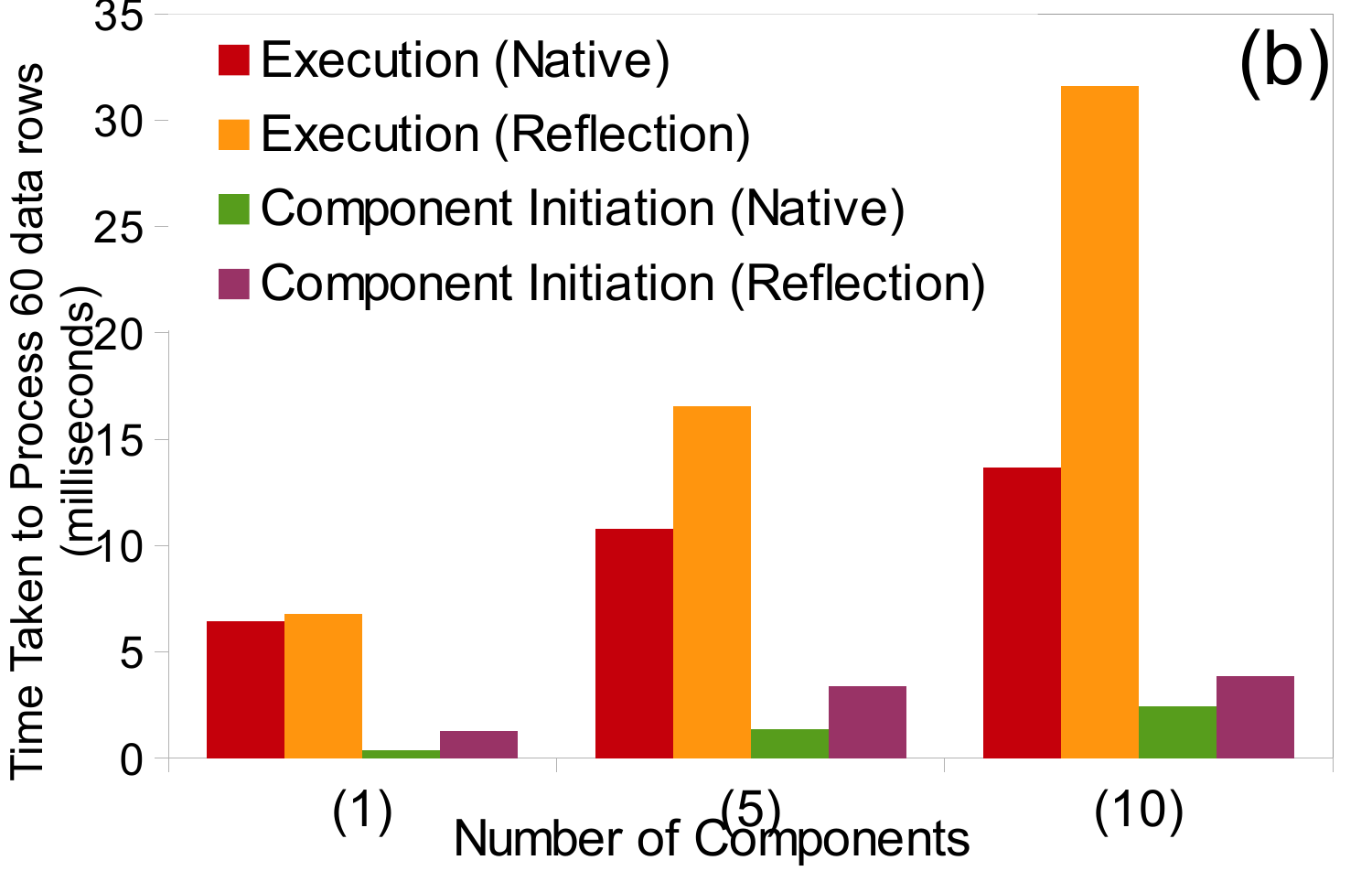} }} }
\mbox{ \subfigure{ \includegraphics[scale=0.27]{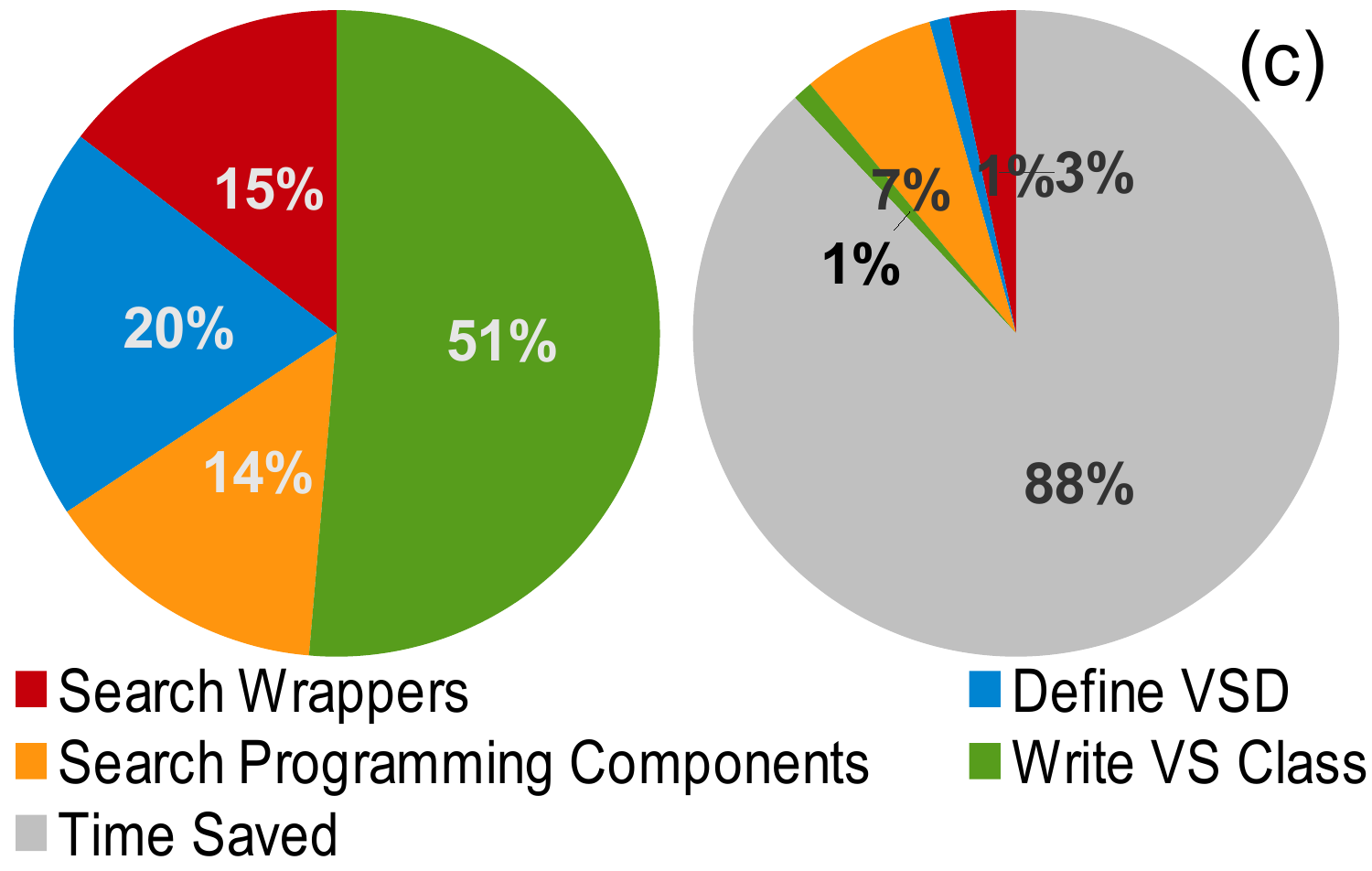}
\quad
\subfigure{ \includegraphics[scale=0.27]{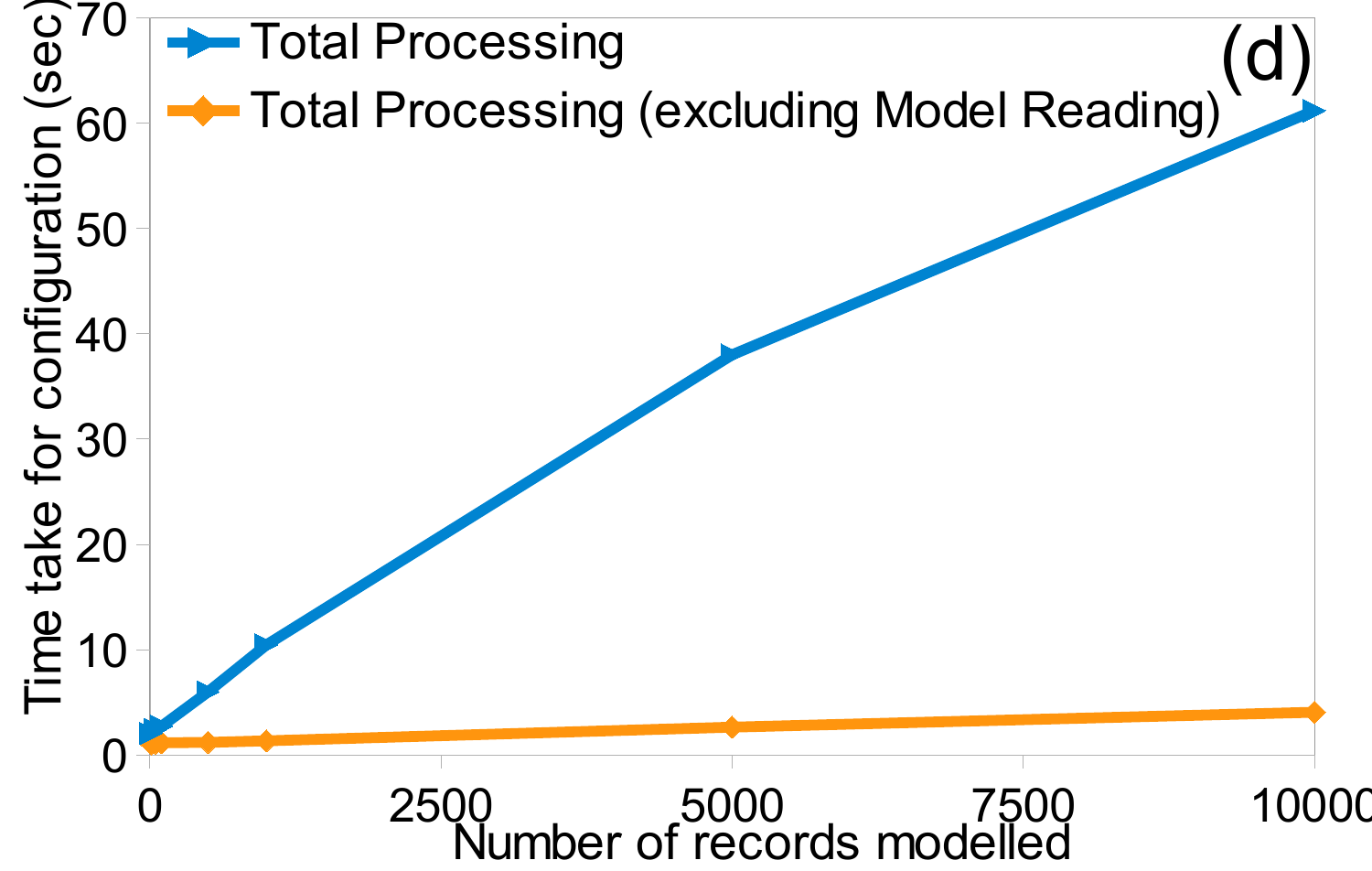} }}}
\vspace{-22pt}
\caption{Evaluation of CASCoM}
\vspace{-20pt}
\label{Figure:Results}
\end{figure}

We evaluated CASCoM in both qualitative and quantitative means. We analysed and compared our proposed solution with respect to the existing GSN configuration model briefly in Figure \ref{Figure:configuration_Workflow_Comparison}. In order to quantify the differences between the two approaches, we evaluated three use case scenarios. In each use case, a user required to configure the IoT middleware in such a way that it produces a specific data stream: (1) \textit{monitor Phytophtora disease}, (2) \textit{monitor environmental pollution}, and (3) \textit{monitor and analyse crowd movement (indoor)}. Further, we selected three types of users: (1) \textit{an IT expert who was familiar with GSN configuration process}, (2) \textit{an IT expert who was not familiar with the GSN}, and (3) an non-IT expert (from medical field). For each use case, a set of basic instructions and programming guidelines that explains the GSN configuration process were given. First, we asked the users to configure the GSN middleware without the support of CASCoM. Secondly, we asked the users to configure the GSN middleware by using CASCoM. We measured time taken by each user and results are presented in Figure \ref{Figure:Results}(a).
 
 In Figure \ref{Figure:Results}(b), we measured how much time it took to process data in a VS using two approaches: (1) to write a customized VS class autonomously by CASCoM and compile it at runtime based on the user requirements (2) to use Java reflection (to support OSGi) so there is no compilation required. Even though the results are obvious, it is important to examine the differences closely, because every data item that comes into GSN will need to go through some VS for  processing. The approach we select has direct impact on the scalability of the GSN as it is expected to retrieve data from a large number of sensors in real world deployments. We employed rule-based reasoning  modules which take some inputs and produce single output (e.g. air stress detection). Then, we increased the number of reasoning operations performed over a single data row. We also measured initiation and execution time separately.

In Figure \ref{Figure:Results}(c), we analysed different phases of the configuration process separately  and compared the current approaches with the CASCoM approach. In order to make the results comparable, we assumed the users are IT experts who know the GSN configuration process. Finally, in Figure \ref{Figure:Results}(d), we added more data into the ontology based semantic models and evaluated the performance of CASCoM by measuring the total execution time. We inserted data to the model by describing more sensors, data processing components, and QA knowledge (e.g. 1000 data records means 1000 sensor descriptions and 1000 data components descriptions and so on). This figure is based on synthetically generated semantic descriptions.

\textbf{Results:} Figure \ref{Figure:Results}(a) shows that CASCoM allows to considerably reduce the time required for configuration of data processing mechanism in IoT middleware. Specifically, CASCoM allowed the three types of users to complete the given task 50, 80 and 250 times faster (respectively) in comparison to the existing approach. \textcolor{black}{According to Figure \ref{Figure:Results}(b), the Java reflection approach takes slightly more time to specially when initializing. Though the Java reflection approach can add more flexibility to our model, the additional overhead increases when the number of components and operation involved gets increased. The overheads can grow up to an acceptable level very quickly when GSN scales up (e.g. more user requests).}

According to Figure \ref{Figure:Results}(c), even  IT experts who know GSN can save time by using CASCoM up to 88\%. Specially, time taken for defining the VSD and VS class have been significantly reduced. Both files can be generated by CASCoM autonomously within a second even for complex scenarios. However, the time taken to find data processing components and sensors (and wrappers) depends on the size of the semantic data model. Figure \ref{Figure:Results}(d) shows how total processing time would vary depending on the size of the semantic data model. Approximately, a semantic model with 10,000 sensor descriptions and 10,000 data processing components can be processed in order to find solutions for a given user request in less than a minute. However, most of the time is taken to read the data model. The actual configuration process other than reading the data model takes only 4 seconds. The actual processing time slightly increases when the number of sensors and data processing components that are described in the model gets increased. In contrast, time required to read the data model increases significantly. However, reading the model is not a frequent repetitive activity and do not make any impact on scalability. In addition, it is less likely that we are required to store very large amounts of data in ontological models in a single GSN instance. Instead, GSN supports distributed processing (i.e. different GSN instances) where CASCoM can collectively process large data sets.

\textbf{Discussion and Lessons Learnt:} Non-IT experts required an extremely detailed guidelines (compared to IT experts) to perform the configuration as there are not familiar with the activities such as programming. They also required direct verbal assistant from the authors. In addition, it was revealed that non-IT experts and IT experts who are not familiar with GSN were unable to configure the GSN at all without guidelines. In contrast, simple guidelines that explain the GUI allowed all users to complete the given task within a fairly similar amount of time. Though the complexity of the user requirement makes visible impact on configuration time in the current GSN approach, it diminishes when users use CASCoM to configure GSN.

VS are designed to be used for data pre-processing. In order to test the capabilities of CASCOM, we also embed context discovery functionalities into the VS. In order to process data, we use a Java based components. We composed data processing components together by creating and compiling a new virtual sensor class. This process is similar to generate a HTML file dynamically when a user requests it (via browser). As a result, at runtime, data is processed natively. We used this approach instead of utilizing  an approach such as web services  due to the fact that IoT middleware needs to deal with real-time data processing. The delay created by additional overheads in web service calling creates bottlenecks to the entire data processing process. We also refrained from using techniques such as Java reflection due to latency and additional overheads. Though the differences are small when considering a single VS as in Figure \ref{Figure:Results}(b), large number of virtual sensors  make significant impact on scalability of the GSN.

As CASCoM stores knowledge in ontologies, users do not need to memorise domain knowledge (i.e. \textit{which sensor data types are required to perform a  certain task?}). This is an significant improvement over the existing approach. Due to the employment of semantic technologies, CASCoM is extensible into any domain. More importantly, adding new sensor descriptions and data processing component descriptions to the data model overtime allows CASCoM to compose new solutions. Ontological reasoning allows to deal with inconsistent usage of domain specific terminologies among domain experts. Ontologies helped in CASCoM to deal with performing validating task in composition of data components. Alternative to ontologies, we could have used a configuration file that explains which programming components and sensors need to be used to produce the required data stream for a given application (e.g. template-base approach). However, such an approach will drastically reduce the interoperability and flexibility. In IoT, ideal approaches should be able to dynamically compose and configure sensors and data processing components as it is impossible predict their availability at give time (new sensors and data processing components may available to use).

\section{Related Work}
\label{sec:Related_Work}

Our solution combines technologies from different research areas such as IoT middleware, semantic technologies, software component composition, and context-aware computing. Microsoft \textit{SensorMap} \cite{P578} (sensormap.org) is a data sharing and visualization framework. It is a peer produced sensor network that consists of sensors deployed by contributors around the world. \textit{SensorMap} mashes up sensor data on a map interface. Then, it allows to selectively query sensors and visualize data. Our approach completely automates the configuration process by eliminating the requirement of hand picking sensors. \textit{Linked Sensor Middleware} (LSM) \cite{P584} (lsm.deri.ie) is a platform that provides wrappers for real time data collection and publishing. It also provides a web interface for sensor search, linked stream data query, data annotation and visualisation. \textit{LSM} mainly focuses on linked data publishing. Sensor selection needs to be done manually in order to retrieve sensor data. \textit{Cosm} (formerly Pachube) (cosm.com) is a platform for Internet of Things devices. \textit{Cosm} allows different data sources to be connected to it. Then, it provides functionalities such as event triggering and data filtering. It acts as a mediator between sensors and applications where users need to manually select and configure sensors.

Context-awareness is a critical functionality that needs to be embedded into IoT middleware solutions \cite{ZMP007}. Context information (e.g. accuracy, reliability, cost) plays a significant role in selecting sensors and data processing components \cite{ZMP006}. To support this, CASCoM provides context discovery functionalities by using semantic knowledge and fusing raw sensor data. The \textit{SensorMashup} \cite{P068} platform offers a visual composer for sensor data streams. Data sources and intermediate analytical tools are described by reference to an ontology, enabling an integrated discovery mechanism for such sources. Selection of data sources and analytical tools based on user requirement need to be done manually by users. Khemakhem et al. \cite{P613} use multiple ontologies to discover and compose software components by focusing on non-functional proprieties. Web service (WS) composition using ontologies \cite{P601} is similar to software component composition performed in CASCoM with a functional point of view but significantly different in implementation and execution point of view. Software component composition is much simpler compared to WS composition \cite{P600} due to lack of overheads. Our solution employs simple data processing components which perform single tasks. We eliminate the burden of interoperability of software components, so they are designed to be used in GSN according to a given specification and stored locally. Therefore, it is not required to handle network communication and complex data structures. However, software components are allowed to call external web services though such actions are not encouraged due to latency that may effect real-time processing. Several projects \cite{P612} have designed and developed ontologies to describe software components. Such approaches have helped them to perform dynamic composition of software components. A process of software component matching using ontologies has been explained in \cite{P605}. In our work we employed the Software Component Ontology discussed in \cite{P612}. Semantic Sensor Ontology (SSNO) \cite{P626} also allowed us to model sensor descriptions. Noguchi et al. \cite{P615} have proposed a mechanism that generates connection between different software components in order to process sensor data and detect events. In contrast, our objective is to produce the data streams required by the users so they can be further analysed extensively using sophisticated applications.

\section{Conclusion and Future Work}
\label{sec:Conclusion}

In this paper, we have introduced the CASCoM approach that allows non-IT experts  to configure IoT middleware efficiently and effectively. The semantic technologies allow to capture user requirements and configure the sensors and data processing components accordingly by handling the low-level technical details without overwhelming the users. Our model supports \textit{single-click} configuration by eliminating sequences of manual activities needed to be carried out by users otherwise. CASCoM makes the configuration process much easier by providing a sophisticate graphical user interface to express user requirements. In addition, our model has the capability to advise users on future sensor deployments in situations where user requirements cannot be satisfied using existing resources. CASCoM also allows to discover additional context information. Users are not required to know any underling technical details. Instead they are offered an user interface where they may select additional context. We propose a cost model that calculates the cost of data acquisition based on sensors and data processing components combined. CASCoM selects the most optimized solution by default, though it allows advance customization through context prioritization. We integrated our model into an IoT middleware called \textit{Global Sensor Networks}. CASCoM has significantly increased the usability and capability of the GSN middleware.

We have shown that it is possible to offer a sophisticated configuration model to support non-IT experts. Semantic technologies are used extensively to support this model. We used ontologies to model sensor descriptions and data processing component descriptions. We also developed a ontology to organize additional knowledge that is required for understanding user requirements. Using our proof of concept implementation, both IT and non-IT experts were able to configure the GSN in significantly less time. In future, we plan to extend our configuration model into sensor-level. To achieve this, we will develop a model that can be used to configure sensors autonomously without human intervention in highly dynamic smart environments in the Internet of Things paradigm. Our approach will explore and identify sensors that are deployed across a given environment autonomously. Future plans include amalgamation of both sensor-level configuration with system-level configuration. Such complete solution will make sensor deployments much faster and easier. CASCoM will stimulate the adaptation of IoT among non technical users due to improved usability.

\subsubsection*{Acknowledgments.} 
This work is supported by European Commission under seventh framework program, contract number FP7-ICT-2011-7-287305-OpenIoT.

\bibliographystyle{IEEEtran}

\bibliography{Bibliography}

\begin{thebibliography}{10}
\providecommand{\url}[1]{#1}
\csname url@rmstyle\endcsname
\providecommand{\newblock}{\relax}
\providecommand{\bibinfo}[2]{#2}
\providecommand\BIBentrySTDinterwordspacing{\spaceskip=0pt\relax}
\providecommand\BIBentryALTinterwordstretchfactor{4}
\providecommand\BIBentryALTinterwordspacing{\spaceskip=\fontdimen2\font plus
\BIBentryALTinterwordstretchfactor\fontdimen3\font minus
  \fontdimen4\font\relax}
\providecommand\BIBforeignlanguage[2]{{%
\expandafter\ifx\csname l@#1\endcsname\relax
\typeout{** WARNING: IEEEtran.bst: No hyphenation pattern has been}%
\typeout{** loaded for the language `#1'. Using the pattern for}%
\typeout{** the default language instead.}%
\else
\language=\csname l@#1\endcsname
\fi
#2}}

\bibitem{P029}
H.~Sundmaeker, P.~Guillemin, P.~Friess, and S.~Woelffle, ``Vision and
  challenges for realising the internet of things,'' European Commission
  Information Society and Media, Tech. Rep., March 2010,
  \url{http://www.internet-of-things-research.eu/pdf/IoT_Clusterbook_March_2010.pdf}
  [Accessed on: 2011-10-10].

\bibitem{ZMP007}
C.~Perera, A.~Zaslavsky, P.~Christen, and D.~Georgakopoulos, ``Context aware
  computing for the internet of things: A survey,'' \emph{Communications
  Surveys Tutorials, IEEE}, vol.~xx, pp. x--x, 2013.

\bibitem{P022}
\BIBentryALTinterwordspacing
K.~Aberer, M.~Hauswirth, and A.~Salehi, ``Infrastructure for data processing in
  large-scale interconnected sensor networks,'' in \emph{International
  Conference on Mobile Data Management}, May 2007, pp. 198--205. [Online].
  Available: \url{http://dx.doi.org/10.1109/MDM.2007.36}
\BIBentrySTDinterwordspacing

\bibitem{ZMP003}
A.~Zaslavsky, C.~Perera, and D.~Georgakopoulos, ``Sensing as a service and big
  data,'' in \emph{International Conference on Advances in Cloud Computing
  (ACC-2012)}, Bangalore, India, July 2012, pp. 21--29.

\bibitem{ZMP004}
\BIBentryALTinterwordspacing
C.~Perera, A.~Zaslavsky, P.~Christen, and D.~Georgakopoulos, ``Ca4iot: Context
  awareness for internet of things,'' in \emph{IEEE International Conference on
  Conference on Internet of Things (iThing)}, Besançon, France, November 2012,
  pp. 775--782. [Online]. Available:
  \url{http://dx.doi.org/10.1109/GreenCom.2012.128}
\BIBentrySTDinterwordspacing

\bibitem{P452}
A.~Baggio, ``Wireless sensor networks in precision agriculture,'' Delft
  University of Technology – The Netherlands, Tech. Rep., 2009,
  \url{http://www.sics.se/realwsn05/papers/baggio05wireless.pdf} [Accessed on:
  2012-05-10].

\bibitem{P598}
\BIBentryALTinterwordspacing
K.~Taylor and L.~Leidinger, ``Ontology-driven complex event processing in
  heterogeneous sensor networks,'' in \emph{Proceedings of the 8th extended
  semantic web conference on The semanic web: research and applications -
  Volume Part II}, ser. ESWC'11.\hskip 1em plus 0.5em minus 0.4em\relax Berlin,
  Heidelberg: Springer-Verlag, 2011, pp. 285--299. [Online]. Available:
  \url{http://dl.acm.org/citation.cfm?id=2017936.2017959}
\BIBentrySTDinterwordspacing

\bibitem{P612}
F.~E. Castillo-Barrera, R.~C.~M. Ram\'{\i}rez, and H.~A. Duran-Limon,
  ``Knowledge capitalization in a component-based software factory: a semantic
  viewpoint,'' in \emph{LA-NMR}, 2011, pp. 105--114.

\bibitem{P626}
M.~Compton, P.~Barnaghi, L.~Bermudez, R.~García-Castro, O.~Corcho, S.~Cox,
  J.~Graybeal, M.~Hauswirth, C.~Henson, A.~Herzog, V.~Huang, K.~Janowicz, W.~D.
  Kelsey, D.~L. Phuoc, L.~Lefort, M.~Leggieri, H.~Neuhaus, A.~Nikolov, K.~Page,
  A.~Passant, A.~Sheth, and K.~Taylor, ``The ssn ontology of the w3c semantic
  sensor network incubator group,'' \emph{Web Semantics: Science, Services and
  Agents on the World Wide Web}, vol.~17, no.~0, pp. 25 -- 32, 2012.

\bibitem{ZMP006}
C.~Perera, A.~Zaslavsky, P.~Christen, M.~Compton, and D.~Georgakopoulos,
  ``Context-aware sensor search, selection and ranking model for internet of
  things middleware,'' in \emph{IEEE 14th International Conference on Mobile
  Data Management (MDM)}, Milan, Italy, June 2013.

\bibitem{P578}
\BIBentryALTinterwordspacing
S.~Nath, J.~Liu, and F.~Zhao, ``Sensormap for wide-area sensor webs,''
  \emph{Computer}, vol.~40, no.~7, pp. 90--93, July 2007. [Online]. Available:
  \url{http://dx.doi.org/10.1109/MC.2007.250}
\BIBentrySTDinterwordspacing

\bibitem{P584}
D.~L. Phuoc, H.~N.~M. Quoc, J.~X. Parreira, and M.~Hauswirth, ``The linked
  sensor middleware - connecting the real world and the semantic web,'' in
  \emph{International Semantic Web Conference (ISWC)}, October 2011.

\bibitem{P068}
\BIBentryALTinterwordspacing
D.~L. Phuoc and M.~Hauswirth, ``Linked open data in sensor data mashups,'' in
  \emph{2nd International Workshop on Semantic Sensor Networks (SSN09)}, 2009,
  pp. 1--16. [Online]. Available: \url{http://ceur-ws.org/Vol-522/p3.pdf}
\BIBentrySTDinterwordspacing

\bibitem{P613}
S.~Khemakhem, K.~Drira, and M.~Jmaiel, ``Semantic matching to achieve software
  component discovery and composition,'' Laboratory for Analysis and
  Architecture of Systems, Tech. Rep., December 2012,
  \url{http://hal.archives-ouvertes.fr/docs/00/79/62/46/PDF/paper12.pdf}
  [Accessed on: 2013-02-05 ].

\bibitem{P601}
E.~Maximilien and M.~Singh, ``A framework and ontology for dynamic web services
  selection,'' \emph{Internet Computing, IEEE}, vol.~8, no.~5, pp. 84--93,
  2004.

\bibitem{P600}
N.~Milanovic and M.~Malek, ``Current solutions for web service composition,''
  \emph{Internet Computing, IEEE}, vol.~8, no.~6, pp. 51--59, 2004.

\bibitem{P605}
C.~Pahl, ``An ontology for software component matching,'' in \emph{Proceedings
  of the 6th international conference on Fundamental approaches to software
  engineering}, ser. FASE'03.\hskip 1em plus 0.5em minus 0.4em\relax Berlin,
  Heidelberg: Springer-Verlag, 2003, pp. 6--21.

\bibitem{P615}
H.~Noguchi, T.~Mori, and T.~Sato, ``Automatic generation and connection of
  program components based on rdf sensor description in network middleware,''
  in \emph{IEEE International Conference on Intelligent Robots and Systems},
  2006, pp. 08--14.

\end{thebibliography}

\end{document}